\documentclass[10pt,preprint]{aastex}
\usepackage[dvips]{color}
\usepackage{graphicx}

\def\msol{\hbox{\kern 0.20em $M_\odot$}}
\def\lsol{\hbox{\kern 0.20em $L_\odot$}}
\def\rsol{\hbox{\kern 0.20em $R_\odot$}}
\def\sr{\hbox{\kern 0.20em sr}}
\def\srmu{\hbox{\kern 0.20em sr$^{-1}$}}
 
\def\g{\hbox{\kern 0.20em g}}
\def\gmu{\hbox{\kern 0.20em g$^{-1}$}}
\def\kg{\hbox{\kern 0.20em kg}}
\def\pc{\hbox{\kern 0.20em pc}}
 
\def\mum{\hbox{\kern 0.20em $\mu$m}}
\def\mumd{\hbox{\kern 0.20em $\mu$m$^{-2}$}}
\def\cm{\hbox{\kern 0.20em cm}}
\def\m{\hbox{\kern 0.20em m}}
\def\km{\hbox{\kern 0.20em km}}
\def\nm{\hbox{\kern 0.20em nm}}
 
\def\s{\hbox{\kern 0.20em s}}
\def\h{\hbox{\kern 0.20em h}}
\def\sec{\hbox{\kern 0.20em sec}}
\def\min{\hbox {\kern 0.20em min}}
\def\smu{\hbox{\kern 0.20em s$^{-1}$}}
\def\smd{\hbox{\kern 0.20em s$^{-2}$}}
\def\an{\hbox{\kern 0.20em an}}
\def\anmu{\hbox{\kern 0.20em an$^{-1}$}}
\def\deg{\hbox{\kern 0.20em $^{\rm o}$}}
\def\yr{\hbox{\kern 0.20em yr}}
\def\yrmu{\hbox{\kern 0.20em yr$^{-1}$}}
\def\Myr{\hbox{\kern 0.20em Myr}}
\def\Mymu{\hbox{\kern 0.20em Myr$^{-1}$}}
\def\K{\hbox{\kern 0.20em K}}
\def\pcmu{\hbox{\kern 0.20em pc$^{-1}$}}
\def\pcmd{\hbox{\kern 0.20em pc$^{-2}$}}
\def\pcmt{\hbox{\kern 0.20em pc$^{-3}$}}
\def\kms{\hbox{\kern 0.20em km\kern 0.20em s$^{-1}$}}
\def\kmpd{\hbox{\kern 0.20em km$^{2}$}}
\def\kpc{\hbox{\kern 0.20em kpc}}
\def\cms{\hbox{\kern 0.20em cm\kern 0.20em s$^{-1}$}}
\def\erg{\hbox{\kern 0.20em erg}}
\def\ergs{\hbox{\kern 0.20em erg}}
\def\cmpd{\hbox{\kern 0.20em cm$^2$}}
\def\cmmd{\hbox{\kern 0.20em cm$^{-2}$}}
\def\cmms{\hbox{\kern 0.20em cm$^{-6}$}}
\def\cmpt{\hbox{\kern 0.20em cm$^3$}}
\def\cmmt{\hbox{\kern 0.20em cm$^{-3}$}}
\def\mpd{\hbox{\kern 0.20em m$^2$}}
\def\mmd{\hbox{\kern 0.20em m$^{-2}$}}
\def\mpt{\hbox{\kern 0.20em m$^3$}}
\def\mmt{\hbox{\kern 0.20em m$^{-3}$}}
\def\mujy{\hbox{\kern 0.20em $\mu$Jy}}
\def\mjy{\hbox{\kern 0.20em mJy}}
\def\Mj{\hbox{\kern 0.20em MJy}}
\def\jy{\hbox{\kern 0.20em Jy}}
\def\ghz{\hbox{\kern 0.20em GHz}}
\def\srmd{\hbox{\kern 0.20em sr$^{-1}$}}
\def \kms{km~$\rm{s}^{-1}$}

\def \mum{$\mu$m}

\def\G{\hbox{\kern 0.20em G}}

\def\h13cop{\hbox{H$^{13}$CO$^{+}$}}

\def\S+{\hbox{S{\small II}}}

\slugcomment{The Evolving ISM in the Milky Way \& Nearby Galaxies}

\shorttitle{ICM-ISM Effects on the FIR-Radio Correlation}
\shortauthors{Murphy et al.}

\begin{document}

\newcommand{\jfourteen}{\hbox{$J=14\rightarrow 13$}}
 \title{How the Intracluster Medium Affects the Far-Infrared--Radio Correlation within Virgo Cluster Galaxies}

\author{
E.J.~Murphy,\altaffilmark{1,2} J.D.P.~Kenney,\altaffilmark{1} 
G.~Helou,\altaffilmark{3} A.~Chung,\altaffilmark{4} and J.H.~Howell\altaffilmark{2}}
\altaffiltext{1}{\scriptsize Department of Astronomy, Yale University,
  P.O. Box 208101, New Haven, CT 06520-8101}
\altaffiltext{2}{\scriptsize {\it Spitzer} Science Center, MC 314-6,
  California Institute of Technology, Pasadena, CA 91125; emurphy@ipac.caltech.edu}
\altaffiltext{3}{\scriptsize California Institute of Technology, MC
  314-6, Pasadena, CA 91125}
\altaffiltext{4}{\scriptsize Jansky Fellow of the NRAO at the University of Massachusetts, Amherst, MA 01003}

\begin{abstract}
We present a study on the effects of the intracluster medium (ICM) on
the interstellar medium (ISM) of 10 Virgo cluster galaxies 
using {\it Spitzer} far-infrared (FIR) and VLA radio continuum imaging.
Relying on the FIR-radio correlation {\it within} normal galaxies, we use our
infrared data to create model radio maps which we compare to the
observed radio images. 
For 6 of our sample galaxies we find regions along their outer edges
that are highly deficient in the radio compared with our models.
We believe these observations are the signatures of ICM ram pressure. 
For NGC~4522 we find the radio deficit region to lie just exterior to a region of high radio polarization 
and flat radio spectral index, however the total radio continuum in this region does not appear significantly enhanced.
This scenario seems consistent for other galaxies with radio
polarization data in the literature.  
We also find that galaxies having local radio deficits appear to have
enhanced global radio fluxes.  
Our preferred physical picture is that the observed radio deficit
regions arise from the ICM wind sweeping away cosmic-ray (CR) electrons
and the associated magnetic field, 
thereby creating synchrotron tails observed for some of our galaxies. 
CR particles are also re-accelerated by ICM-driven shocklets behind the observed
radio deficit regions which in turn enhances the remaining radio disk brightness.  
The high radio polarization and lack of coincidental signatures
in the total synchrotron power in these regions arises from shear, and possibly mild compression, as the ICM wind drags and stretches the magnetic field. 
\end{abstract}

\lefthead{Murphy et al.}
\righthead{ICM-ISM Effects on the FIR-Radio Correlation}

\section{Introduction}
The physical processes associated with interactions between the
intracluster medium (ICM) and the interstellar medium (ISM) play
critical roles driving the evolution of spiral galaxies in clusters
\citep[e.g.][]{gg72,ltc80,amb99,ss01,bv01}.  
Galaxies are preferentially found in groups or clusters where most of
these processes occur, yet many basic effects related to ICM-ISM 
interactions (i.e. ram pressure stripping) are still not well understood.  
These effects include the fate of star-forming molecular clouds, the
rates of triggered star formation  versus gas removal, and the
possible reconfiguration of a galaxy's large-scale magnetic field
and/or cosmic-ray (CR) particles.  

Since ram pressure will more easily affect lower density constituents
of the ISM, it seems that the diffuse radio continuum halos of
galaxies may be sensitive tracers of active ICM pressure as indicated
by observations for a number of Virgo cluster spirals including 
NGC~4522 \citep{bv04}, NGC~4402 \citep{hc05}, NGC~4254 \citep{kc07}, 
and a few others \citep{bv07,ac07}.  
More specifically, large regions of enhanced polarized radio continuum
emission have been found within a number of cluster spirals
\citep{bv07, kc07}; 
the maxima of the polarized radio continuum distributions within these
galaxies are located along outer edges and thought to arise from
external influences of the cluster environment.  
However, without a good idea of the galaxy's unperturbed appearance in
the radio, these observations alone make it difficult to quantify the
extent of ram pressure effects.
We now address this comparison using the nearly universal correlation
between the far-infrared (FIR) and non-thermal radio continuum
emission of normal galaxies \citep[e.g.][]{de85,gxh85}. 

Using a phenomenological smearing model, in which a galaxy's FIR map is
smoothed by a parameterized kernel to compensate for the fact that the
mean free path of dust heating photons is much shorter than the
diffusion length of CR electron, \citet{ejm08a} has shown for a sample
of 15 non-Virgo spiral galaxies that the dispersion in the FIR/radio ratios
on sub-kiloparsec scales {\it within} galaxies can be reduced by a
factor of $\sim$2, on average.    
Accordingly, it is possible to obtain a good first order approximation
of an undisturbed galaxy's non-thermal radio continuum morphology with
its FIR image alone. 

Coupling FIR observations taken by the {\it Spitzer} Space Telescope with VLA radio continuum imaging, obtained as part of the VLA Imaging of Virgo in Atomic gas (VIVA; A. Chung et al. 2008, in preparation) survey, we study how the relativistic and gaseous phases of the ISM are affected by ICM-ISM  interactions for a sample of Virgo cluster galaxies. 
This is done using FIR {\it Spitzer} maps to predict how the radio morphology should appear if the galaxy resided in the field;
in this paper we make the case that significant deviations from such an appearance are directly related to ICM-ISM interactions. 
See \citet{ejm08b} for the complete study.

\section{Observations and Analysis}
A total of $\sim$40 of the 53 VIVA sample galaxies are included in the {\it Spitzer} Survey of Virgo (SPITSOV; see J.D.P.~Kenney et al. 2008, in preparation; J.D.P~Kenney et al. this proceedings) imaging program.
At the time of this writing, a sub-sample of 10 galaxies had existing high quality FIR and radio continuum maps.
{\it Spitzer} imaging was carried out for each galaxy using the Multiband Imaging Photometer for {\it Spitzer} \citep[MIPS;][]{gr04}. 
The strategy of the imaging campaign was based on that of the Spitzer Infrared Nearby Galaxies Survey \citep[SINGS;][]{rk03} with the only difference being a factor of 2 increase in exposure times to better detect diffuse emission arising in the outer regions of the Virgo sample galaxies.

The 1.4~GHz radio continuum maps were created from the line-free channels of H~{\sc i} data cubes collected as part of of the VIVA survey; 
a detailed description of the VLA observations along with the H~{\sc i} reductions and associated data products can be found in A. Chung et al. (2008, in preparation).  
To perform accurate comparative analysis between the MIPS and radio data we match the resolution of the final calibrated images using the MIPS PSF. 
After cropping each set of galaxy images to a common field of view we CLEANed the radio data and convolved the resulting CLEAN components with a model of the MIPS 70~$\micron$ PSF.

\subsection{Modeled Radio Continuum Maps \label{sec-modrc}}
Detailed studies of the FIR-radio correlation within nearby field
galaxies \citep[e.g.][]{ejm06a, ejm08a} have shown that the dispersion
in the FIR/radio ratios on $\ga$$0.1 - 1$~kpc scales within galaxy
disks is similar to the dispersion in the global FIR-radio correlation
(i.e. $\la$0.3~dex).  
This strong correlation between the FIR and radio images of spirals has
been found to improve by a factor of $\sim$2 using an image-smearing
technique which approximates a galaxy radio image as a smoother version
of its FIR image due to the diffusion of CR electrons.
While the star-forming disks of cluster galaxies are often truncated
due to the gas stripping events, the 70~$\micron$ images for the
entire sample do not appear to be more strongly disturbed than galaxies in
the field; 
we do not observe sharp edges, tails, or clear evidence for extraplanar
emission at 70~$\micron$.
Therefore, we apply the image-smearing technique of \citet{ejm06b,ejm08a}
(using a single smoothing function) to create models for the expected
radio distribution of a galaxy assuming that the FIR/radio ratio map is like that
of a field galaxy.

\begin{figure}[!ht]
\centerline{\hbox{
  \resizebox{15cm}{!}{
\plotone{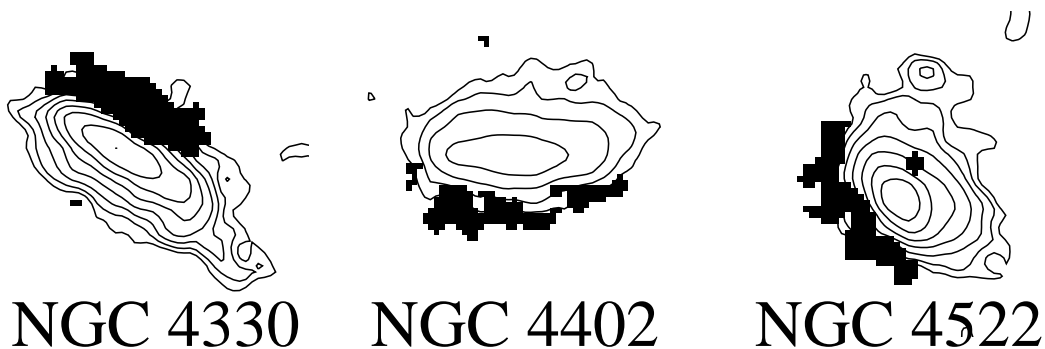}}}}
\caption{\footnotesize
The radio deficit regions of NGC~4330, NGC~4402, and NGC~4522 
with radio continuum contours.  
The radio continuum contours begin at the 3-$\sigma$ RMS level and increase logarithmically.  
For each galaxy we find the deficit region to be located on the edge opposite that of the observed synchrotron tails.  \label{fig-1}}  
\end{figure}

\subsection{Deviations from Expectations for Field Galaxies}
Our aim is to determine if differences between the observed and modeled radio continuum images arise from ICM-ISM interactions and, if so, whether quantifying and comparing such differences within our sample can help improve our current physical picture of such interactions.  
We therefore create ratio maps between the observed and modeled radio maps in the following  manner.
We divide the observed radio map by the modeled radio map after having removed pixels not detected at the 5-$\sigma$ RMS level of the modeled map only. 
Pixel values in the ratio map $\geq$1.3  and $\leq$0.5 are considered to be excesses and deficits, respectively; for a more complete description on how radio deficit and excess regions were determined see \citet{ejm08b}.

For most (60\%) of our sample galaxies we detect radio deficit regions which are located along a single edge of their disks and opposite of any identified H~{\sc i} tails.  
The radio deficit regions are also generally found to be opposite any radio excess regions associated with synchrotron tails; this is illustrated in Figure \ref{fig-1} for NGC~4330, NGC~4402, and NGC~4522.  
Since these H~{\sc i} and synchrotron tails are likely identifying the direction of the ICM wind, we focus our attention toward the radio deficit regions as they are probing directly the most intense effects of the ongoing ICM-ISM interactions.   
The combination of these observations suggests that the radio deficit regions likely arise from the same gravitational or gas dynamical effects which have displaced the galaxy's H~{\sc i} gas and relativistic plasma.  
We therefore believe that the radio deficit regions identify the zone in which the ICM wind is actively working on each galaxy's ISM.

To compare quantitatively the radio deficiencies among the sample galaxies we define the parameter
\begin{equation}
\Upsilon = \frac{(S_{\nu}^{\rm mod} - S_{\nu}^{\rm obs})_{\rm
    def}}{S_{\nu}^{\rm glob}}
\end{equation}
which measures the difference between the observed and modeled radio flux densities within the radio deficit region normalized by the global radio flux density of the galaxy.

\begin{figure}[!ht]
\centerline{\hbox{
  \resizebox{9cm}{!}{
\plotone{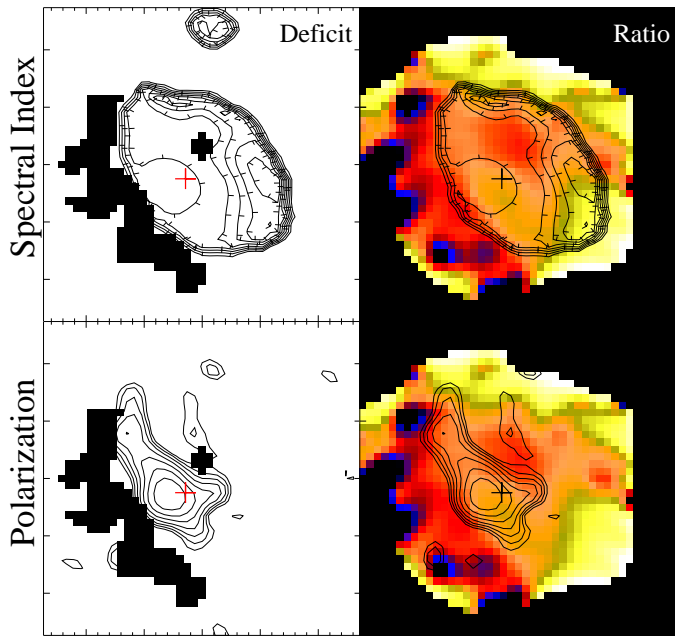}}}}
\caption{\footnotesize
  In the top panels, from left to right, we plot the radio continuum
  deficit map and the ratio map of the observed to modeled radio continuum
  emission of NGC~4522 overlayed with radio spectral index contours
  taken from \citet{bv04}.  
  The contour levels increase by 0.2 and range between $-1.9$ to
  $-0.3$; the dashes indicate the direction of the downward
  gradient.  
  The ratio maps is show in a logarithmic stretch ranging from 10\% (dark) to a factor of 3 (light). 
  We plot the same images in the bottom panels except this time we
  overlay them with polarized radio contours \citep{bv04}.
  The contours are logarithmically scaled and begin at
  the 3-$\sigma$ RMS level.
  The cross in each panel identifies the center of the galaxy.
  \label{fig-2}}
\end{figure}

\section{NGC~4522: Comparison of Radio Deficit Region with Radio Polarization and Spectral Index Maps}
Radio polarization and spectral index data of NGC~4522 provide evidence for an ongoing ICM-ISM interaction \citep{bv04};
the eastern edge of NGC~4522 is found to be highly polarized and coincident with the flattest spectral index which steepens towards the western side of the galaxy.  
In the first column of Figure \ref{fig-2} we plot the radio
continuum deficit regions and overlay the 20 to 6~cm spectral index
and 6~cm polarized radio continuum maps of \citep{bv04}.
We find that the regions of high polarization and flattest spectral
index lie just interior to the radio deficit region.  
While the peak in the polarized radio continuum is coincident with the
flattest part of the spectral index distribution, this is not where
the total radio intensity (or FIR) peaks as pointed out by
\citep{bv04}.   

These observations suggest that any pre-existing gradient in the
observed FIR/radio ratio distribution will be stronger than deviations
arising from the effects of ram pressure and that, to first order,
star formation in the disk drives the appearance of the FIR/radio
ratios.   
However, by inspecting the ratio map of NGC~4522 displayed in the
second column of Figure \ref{fig-2}, which was created using our
{\it smoothed} 70~$\micron$ image, ICM-ISM effects become more apparent.    
We find slightly higher ratios (i.e. ratios of $\sim$90\%) near the
regions of high radio polarization and flat spectral indices;  
the elevated ratios in these regions are therefore attributed to a
modest increase in the total radio continuum.
We now try to determine the most plausible physical scenario to
explain this combination of observations.

\section{Discussion}
We have shown that a large fraction of our sample of cluster galaxies
exhibits statistically significant radio deficit regions relative to
what the phenomenological image-smearing model of \citet{ejm08a} would
predict within their disks.  
Such deficit regions are not found within normal field galaxies
suggesting that cluster environment processes are likely at work.
Comparing our results with other observational data for NGC~4522,
a galaxy which is clearly experiencing the effects of ram pressure, we
find our deficit region agrees with such a scenario and may be able to
add new insight on the magnitude of the interaction.  

Similar to the example of NGC~4522, radio polarization observations of
other Virgo cluster galaxies exhibit highly polarized radio
emission along the predicted leading edges between their ISM and the 
ICM wind \citep{bv07,kc07}.  
For the galaxies overlapping in our sample which show evidence for
being disturbed (i.e. NGC~4254, NGC~4388 and NGC~4402), the regions of highly polarized continuum emission are coincident with local radio continuum enhancements found in our ratio maps (i.e. ratios of $\sim$$1.1-1.7$) and interior to our radio deficit regions.    
Thus, the scenario described for NGC~4522 seems to be applicable for
these galaxies as well based on the polarized radio continuum results.   
With this picture we will use the findings from our analysis to
quantitatively assess the strength of the ram pressure from the
intracluster medium (ICM).

\subsection{Estimates of Minimum Ram Pressure and Internal
  Relativistic ISM Pressure \label{sec-pram}} 
Ram pressure from the ICM is simply defined as, 
\begin{equation}
\label{eq-pram}
P_{\rm ram} = \rho_{\rm ICM} v_{\rm gal}^{2}
\end{equation}
where $\rho_{\rm ICM}$ is the ICM mass density and $v_{\rm gal}$ is
velocity of a galaxy relative to the ICM.
If the ICM ram pressure exceeds that of a galaxy's relativistic ISM
(CRs $+$ magnetic field) then it should be possible to redistribute
and even strip them from the galaxy disk.  
Using the predicted radio flux density for each deficit region, we can
approximate a minimum value for the ICM ram pressure needed to cause the
observed depression in the radio.  

Taking the predicted flux density of the deficit region along with its
area we use the revised equipartition and minimum energy formulas of
\citet{bk05} to calculate the minimum energy magnetic field strength
of the deficit regions. 
This calculation assumes a proton-to-electron number density ratio of
100, a radio spectral index of $-0.85$, and a path length through the
emitting medium of $1/cos(i)$~kpc where $i$ is the galaxy
inclination. 
Using these magnetic field strengths $B$ (i.e. calculated using
Equation 4 of \citet{bk05}) we compute the magnetic field energy
densities $U_{\rm B} = B^2/(8\pi)$ of the deficit regions.  

Assuming minimum energy between the magnetic field and CR particle
energy densities, $U_{\rm B}$ and $U_{\rm CR}$ respectively, we can
use the values of $U_{\rm B}$ over the deficit region in our model to
determine the minimum $P_{\rm ram}$ necessary to create the deficit
regions.    
From minimum energy arguments we find that $U_{\rm B} = 3/4U_{\rm CR}$. 
For a relativistic gas the magnetic pressure, $P_{\rm B}$, and CR pressure $P_{\rm CR}$ are related to energy density such that $P_{\rm B} = U_{\rm B}$ and $P_{\rm CR} = 1/3U_{\rm CR}$ leading to the relation that $P_{\rm CR} = 4/9 U_{\rm B}$.
Then, the pressure of the relativistic ISM is found to be 
\begin{equation}
\label{eq-PrISM}
P_{\rm RISM} = P_{\rm CR} + P_{\rm B} \sim 13/9 U_{\rm B}.
\end{equation}
These estimates, denoted as $P_{\rm RISM}^{\rm mod}$ set the minimum ICM ram pressure necessary to create the observed radio continuum deficit regions (i.e. $P_{\rm ram} > P_{\rm RISM}^{\rm mod}$).  

All galaxies have $P_{\rm RISM}^{\rm mod}$ values between $\sim$$2 - 4 \times 10^{-12}$~dyn~cm$^{-2}$.
Using the projected linear distances to the cluster center among our sample along with a measured density profile of Virgo \citep{mats00} yields a range in ICM density of  $n_{\rm ICM} = \rho_{\rm ICM}/m_{p} \approx 0.6 - 4\times10^{-4}$~cm$^{-3}$.  
Taking a typical Virgo galaxy velocity of 1500~km~s$^{-1}$
\citep{kvv04} the associated range in ICM ram pressure is $P_{\rm ram}
\approx 2 - 15 \times 10^{-12}$~dyn~cm$^{-2}$.  
This range of ICM pressures is generally greater or similar to the
values for $P_{\rm RISM}^{\rm mod}$ which is consistent with ICM ram
pressure being able to create the observed deficit regions.
We also note that these pressures agree with those produced in the 3D
hydrodynamical simulation of \citet{ermb07} over the same projected
linear distances to the cluster center among our sample.

\begin{figure}[!ht]
\centerline{\hbox{
  \resizebox{9cm}{!}{
\plotone{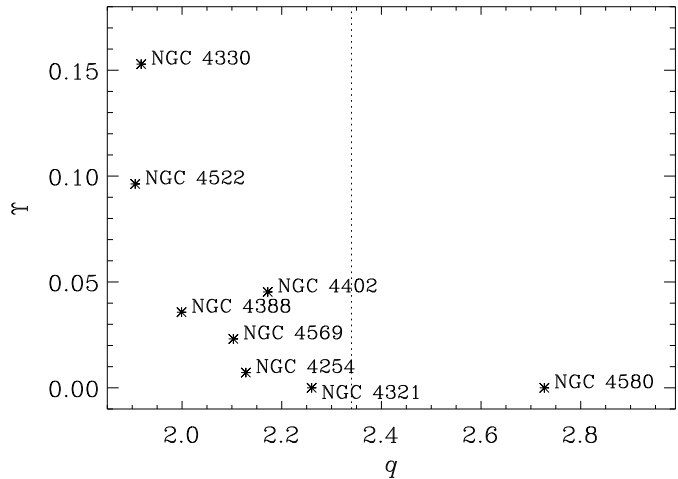}}}}
\caption{\footnotesize
The severity of the deficit region characterized by
  $\Upsilon$ plotted against $q$ (the
  logarithmic FIR/radio ratio). The vertical line at $q = 2.34$ identifies the average $q$ value reported by \citet{yrc01} for 1809 galaxies. \label{fig-3}}
\end{figure}

\subsection{Cosmic Ray Electron Escape and Re-acceleration \label{sec-reacel}} 
While the CR electrons are being moved around by the ICM wind, it does
not appear that they escape the galaxy disks as global $q$
(logarithmic FIR/radio) ratios do not appear systematically high with
respect to the nominal value of $\sim$$2.34 \pm 0.26$~dex
\citep[i.e.][]{yrc01}.  
In fact, by plotting $q$ against $\Upsilon$ (our parameter defining the
severity of the deficit region) in Figure \ref{fig-3} we find a
trend of decreasing FIR/radio ratio with increasing $\Upsilon$
and that nearly all of the $q$ values are lower than the nominal
value.  
Other, more detailed studies on the global FIR and radio properties of
cluster galaxies, have pointed out notably lower FIR/radio ratios than
expected from the FIR-radio relation in the field 
\citep[][]{mo01,ry04}.  
This suggests that there is either depressed FIR emission or an
enhancement in the global radio emission among our sample galaxies.  
And while the trend in Figure \ref{fig-3} may simply be the result
of small number statistics, it also might be suggestive that $q$
itself is sensitive to the strength of the ICM ram pressure and
perhaps a galaxy's stripping history.

The only major exception in our sample is NGC~4580 which has a very high $q$ value; 
this galaxy also has the highest H~{\sc i} deficiency among the sample (a factor of 5 larger than the median).  
It is also thought to have been stripped long ($\sim$400~Myr) ago and is the only post-strongly stripped galaxy observed so long after peak pressure \citep{hc08}.
We therefore speculate that its high $q$ value is the result of CR
electron escape as the ICM wind has swept out most of the galaxy's
gaseous and relativistic ISM long ago. 

While it is still not completely clear, we believe the most plausible explanation for the depressed $q$ values seems to be that of the ICM wind raising the global CR electron energy and synchrotron power by driving shocklets into the ISM which are re-accelerating CR electrons.  
The fact that we only find moderate local enhancements to the total continuum behind the radio deficit region suggests that the shocks run through the entire galaxy disk rather quickly.  
The shock speed, $v_{\rm s}$, which must be super-Alfv\'{e}nic to re-accelerate CR electrons, should have a value around 
\begin{equation}
\label{eq-vshock}
v_{\rm s} \approx \left(\frac{4}{3}\frac{P_{\rm ram}}{\rho_{\rm ISM}}\right)^{1/2} \approx
v_{\rm gal} \left(\frac{4}{3}\frac{\rho_{\rm ICM}}{\rho_{\rm ISM}}\right)^{1/2}, 
\end{equation}
where $P_{\rm ram}$ is the ram pressure, as defined in Equation \ref{eq-pram},
and $\rho_{\rm ISM}$ and $\rho_{\rm ICM}$ are the ISM and ICM densities, respectively.
Taking $v_{\rm gal} \approx 1500$~km~s$^{-1}$, $n_{\rm ISM} = \rho_{\rm ISM}/m_{p} \approx 1$~cm$^{-3}$, and the range in $\rho_{\rm ICM}$ values discussed leads to shock velocities ranging between $\sim$10 to 35~km/s.  
Assuming a thin disk thickness of 500~pc, the shocks should run through each disk on the order of $\sim$15 to 50~Myr; indeed this is very short compared to the dynamical timescale of these systems.

\section{Summary and Conclusions}
We have studied the interstellar medium (ISM) of 10 Virgo galaxies
included in the VLA Imaging of Virgo in Atomic Gas (VIVA) survey using
{\it Spitzer} MIPS and VLA 20~cm imagery.  
By comparing the observed radio continuum images with modeled
distributions, created using a phenomenological image-smearing model
described by \citet{ejm06b, ejm08a}, we find that the edges of many cluster
galaxy disks are significantly radio deficient. 
These radio deficit regions are consistent with being areas affected
by intracluster medium (ICM) induced ram pressure as suggested by
the location of H~{\sc i} and radio continuum tails.
From our results we are able to conclude the following:
\begin{enumerate}
\item
    The distribution of radio/FIR ratios within cluster galaxies thought to be experiencing ICM-ISM effects are systematically different from the distribution in field galaxies;
  radio/FIR ratios are found to be significantly low along galaxy
  edges probably in the direction of the ICM wind arising from a local
  deficit of radio continuum emission.  

\item
  In the case of NGC~4522 we find that the radio deficit region lies
  exterior to a region of high radio polarization and a flat radio
  spectral index.   
  We interpret this to suggest that CR electrons in the halos of
  galaxies are being swept up by the ICM wind.   
  The ICM wind drives shocklets into the ISM of the galaxy which re-accelerate CR particles interior to the working surface at the ICM-ISM interface.
  Some CR electrons may also be redistributed downstream creating
  synchrotron tails as observed for NGC~4522. 
  The high radio polarization is probably the result of shear as
  the ICM wind stretches the magnetic field;
  compression may also play a role, though a modest one, since the
  total radio continuum in these regions does not appear significantly
  enhanced.  
  
\item
   The global radio/FIR ratios of these cluster galaxies are systematically higher than the average value found for field galaxies and appear to increase with increasing severity of the ISM stripping. 
   We attribute this to a greater relative increase in the CR energy density pursuant to a more sever effect of stripping on the galaxy.

\item
  Using the identified radio deficit regions we are able to get a
  quantitative estimate of the minimum strength of the ICM pressure
  required to affect a galaxy disk; 
  we find values in the range of $\sim$$2-4\times
  10^{-12}$~dyn~cm$^{-2}$.  
  These pressures are generally smaller than, but similar to, those
  estimated for typical values of the ICM gas density and galaxy
  velocities, as well as the range of ram pressures calculated by 3D
  hydrodynamical simulations;
  therefore, our estimates are consistent with scenario of
  ram pressure creating the observed deficit regions.

\end{enumerate}

\acknowledgements
We are grateful to the SINGS team for producing high quality data sets
used in this study.  
This work is based in part on observations made with the {\it Spitzer}
Space Telescope, which is operated by the Jet Propulsion Laboratory,
California Institute of Technology under a contract with NASA. 
Support for this work was provided by NASA through an award issued by
JPL/Caltech.  


\end{document}